\documentclass[aps,prl,twocolumn,showpacs,superscriptaddress,groupedaddress,nofootinbib]{revtex4-1}  
\usepackage{graphicx}  
\usepackage{dcolumn}   
\usepackage{bm}        
\usepackage{amssymb}   
\usepackage{amsmath}
\usepackage{xspace}
\usepackage{epstopdf}
\usepackage{wrapfig}
\usepackage{color}
\usepackage{bbm}
\usepackage{color}

\usepackage{ulem} 
\usepackage{subfigure}

\usepackage{times}
\usepackage{color}
\usepackage{ulem}

\newcommand{\s}{$S$}
\newcommand{\sbar}{$\bar{S}$}

\newcommand{\be}{\begin{equation}}
\newcommand{\ee}{\end{equation}}

\begin{document} 
\title{Stable Sexaquark} 


\author
{Glennys R. Farrar }
\affiliation{Center for Cosmology and Particle Physics,
Department of Physics,
New York University, NY, NY 10003, USA}


\begin{abstract}  
It is proposed that the neutral, B=2, flavor singlet sexaquark (\s) composed of $uuddss$ quarks, has mass $m_S \lesssim 2$ GeV.  If $m_S < 2 (m_p + m_e)$, it is absolutely stable, while for $m_S < m_p+m_e + m_\Lambda$, $\tau_S$ can be $> \tau_{\rm Univ}$.  Lattice gauge theory cannot yet predict $m_S$ but indirect evidence supports the hypothesis of stability.   A stable \s\ is consistent with QCD theory and would have eluded detection in accelerator and non-accelerator experiments.  If it exists, the \s\ is a good Dark Matter candidate.  Analyses of existing Upsilon decay and LHC data are proposed which could discover it and measure its mass. 

\end{abstract}
\maketitle

\section{Introduction}
\label{sec:intro}

Fermi statistics gives the 6-quark combination 
$uuddss$ a privileged status.  Uniquely among 6-light-quark states, the spatial wavefunction of the \s\ can be totally symmetric, at the same time that the color, flavor, and spin wavefunctions are simultaneously individually totally antisymmetric.   (Other 6-quark states are not spatially symmetric, e.g., the deuteron is a loosely bound pair of nucleons, and states containing heavier quarks would not be stable.)  The \s\  is 
a spin-0, flavor-singlet, parity-even boson with Q=0, B=2 and S=-2.  Very general most-attractive-channel arguments, e.g., \cite{RDSTumbling80,PeskinVacAlign80,PreskillMAC81}, 
imply the \s\  should be the most-tightly bound state of its class, analogous to the He nucleus' deep binding and the singlet hyperfine state in hydrogen being lower in energy than the triplet.  

The tendency for attraction in the $uuddss$ case was pointed out by R. Jaffe 40 years ago.  He used a 1-gluon-exchange bag model calculation of hyperfine splitting \cite{jaffe:H} to estimate a mass of $\approx 2150$ MeV and called the state the H-dibaryon.   With this mass, $m_H > m_p+m_e + m_\Lambda$ and the H has a typical weak interaction lifetime $\mathcal{O}(10^{-10}$s).  
Numerous experiments to discover the H-dibaryon had null results.   Observation of hyperon decay products from an S=-2 hypernucleus, was widely taken to be evidence against the existence of a strong-interaction stable H-dibaryon \cite{ahn+hypernucBNL01,takahashi+hypernucKEK01}.   
  Strong limits on production of a narrow $\Lambda p \pi^-$ resonance in the final states of $\Upsilon$ decay \cite{BelleH13}, and negative searches for new neutral decaying particles \cite{NA3longLiveNeuts86,bernstein+LongLivedNeut88,belz+96,KTeVHdecay99}, are evidence against an H-dibaryon with a typical weak-interaction lifetime.   The beam-dump + time-of-flight experiment to search for new long-lived neutral particles of Gustafson et al. in 1976 \cite{gustafson} explicitly excluded consideration of masses less than 2 GeV due to the overwhelming neutron background. \cite{suppMat}

However instead of being fairly loosely-bound, the singlet $uuddss$ sexaquark may be a deeply bound state with low enough mass to be stable or essentially stable.  It is shown below that such a particle could have eluded all searches to date.  Two experiments are proposed to discover it if it exists.  The particle is denoted ``$S$" for Sexaquark, Singlet, Scalar, Strong and Stable; a fresh designation is appropriate because the proposed tightly bound \s\ bears scant resemblance to Jaffe's short-lived dihyperon bound state mostly considered before.  

Eventually lattice QCD should be able to answer the question of whether a stable sexaquark exists.  The best study to date shows a strongly bound state in the B=2, S=-2 channel \cite{Beane+13}, but that study is still far from the physical limit with respect to quark masses and infinite volume.  It uses quark masses of 850 MeV, so the quarks are non-relativistic which is far from the  realistic ultra-relativistic situation.  Theory provides a guide as to how to extrapolate meson and baryon masses calculated for feasible quark masses, to the physical few-MeV quark mass limit; such guidance is not available for the \s, whose behavior in the relativistic-quark regime must be determined by careful simulations.  Moreover reaching the infinite volume limit is exponentially more difficult with increasing number of constituents \cite{LepageTASI}.   

Baryon number conservation implies that the \s\ is absolutely stable if $m_S  \le 2 \, (m_p + m_e) = 1877.6$ MeV, while  
if $m_S < m_p+m_e + m_\Lambda = 2054.5 $ MeV, it decays via a doubly-weak interaction and its lifetime could be longer than the age of the Universe \cite{fzNuc03}.  I will drop the distinction and call both cases ``stable" below. 

The general argument for strong attraction in the flavor-spin-color singlet channel, which is supported by lattice QCD  \cite{Beane+13}, implies there should be a $uuddss$ bound state with mass $< 2 m_\Lambda$, but experiment disfavors such a state with mass $\gtrsim 2\,$GeV.  Together, this is indirect evidence for a stable state with mass $\lesssim$ 2 GeV.

A stable \s\  is a potentially excellent dark matter candidate \cite{fDMtoB18,fSDM_ICRC17}, a possibility first noted in \cite{fArkady03}.  The predicted value of $\Omega_{DM}/\Omega_b$ for sexaquark DM, which follows from the Boltzmann equation in the quark-gluon plasma in an essentially model-independent way, agrees well with observation \cite{fDMtoB18}.
Existing direct detection limits are in general inapplicable due to  \s\ scattering with gas in the Galactic disk, which can drive SDM toward approximate co-rotation with the solar system \cite{wfHIDM18}.  (Direct detection limits were improved in \cite{mfWindow17} but without including co-rotation.  Standard limits have recently been shown to require further updating due to inefficient thermalization of the signal in XQC \cite{mfVelDep18} and non-trivial quantum mechanics in the $A$-dependence of the cross-section \cite{fxHIDM18}.)
The mild heating of the gas induced by DM-gas interactions can help resolve the star-formation-quenching problem of LCDM\cite{wfHIDM18} and X-ray cluster cooling \cite{QinWu01,chuzNusser06,wfHIDM18}.   Primordial nucleosynthesis limits on unseen baryons do not apply because \s's are not attracted to nucleons and do not form nuclei \cite{fgBBN18}.  
To first approximation the formation of structure proceeds as for LCDM, but \s-nucleon scattering exerts a very tiny drag which may have observable effects \cite{fSDM_ICRC17}; it is not yet clear whether this may be a source of tension \cite{dvorkin+14,gluscevic+17}, or in fact helps alleviate existing tensions within LCDM such as local versus CMB-derived values of $H_0$ and $\sigma_8$ \cite{L+S15}.  
\vspace{-0.04in} 
\section{Mass of the sexaquark}
\vspace{-0.04in}  
There is no good empirical analog for estimating $m_S$  based on other hadron masses.  The \s\ is a scalar, so chiral symmetry breaking has no obvious implications for its mass, and the mass  of the \s\ has no \textit{a priori} relation to the masses of baryons.   Model predictions for the H-dibaryon mass cover a wide range from stable to unbound.  Kochelev calculated a mass of 1718 MeV in the instanton liquid model \cite{kochelev99} (citing \cite{dorokhovKochelevZubov92} in disputing the claim of  \cite{takeuchiOka90} that three-body repulsion due to light-quark-instanton coupling unbinds the H entirely); an independent assessment is needed.   

A stable 6-quark state seems implausible to most particle physicists, due to the widespread but unjustified reliance on intuition from the naive constituent quark model: the prescription that each valence light (strange) quark has an effective mass of $\sim 300 \, (450)$ MeV.  
However the non-relativistic, weakly-bound constituent quark picture is not theoretically justified at all.  The light quark masses are $<5$ MeV, and the strange quark mass is $<100 $ MeV (c.f. {\cite{pdg16}).  Hence hadrons made of light and strange quarks are highly relativistic bound states, whose mass is virtually entirely dynamical.   

The invalidity of the naive quark model is manifest in the masses of the pseudoscalar mesons: $\pi^{0,\pm}$ masses are $135$ and $140$ MeV instead of $\sim 600$ MeV, and the $\eta'$ mass is 958 MeV rather than $\sim 600-700$ MeV.  The low masses of octet pseudoscalar $q \bar{q}$ mesons are associated with their being the pseudo-Goldstone bosons of chiral symmetry breaking, while the large mass of the $\eta'$ relative to the octet pseudoscalars is attributed to the chiral anomaly \cite{Witten79,Veneziano79}.   But from the perspective of relativistic quark-gluon bound states, the diversity of masses of hadrons containing light ($u,d,s$) quarks arises from  the subtleties of relativistic dynamics in different flavor and spin combinations of quarks, anti-quarks and gluons, and the existence of fundamentally non-perturbative gluon and quark condensates.  

Only 16\% binding energy relative to two $\Lambda$'s is enough for the \s\ to be absolutely stable, or 10\% to have a lifetime longer than the age of the Universe\cite{fzNuc03} -- small compared to the $\mathcal{O}(1)$ mass shifts relative to the naive quark model in the meson sector.  This, along with the indirect argument that some binding in the singlet channel is practically inescapable, yet a mass above 2 GeV appears excluded experimentally, is strong motivation for experimental searches for a stable \s.  

\section{Properties, Interactions and Stability} 


Three crucial attributes of the conjectured stable \s\ are responsible for its not having been detected so far:\\ 
$\bullet$ Its mass makes it difficult to distinguish kinematically from the neutron. \\
$\bullet$ It is neutral and a flavor-singlet, so it does not couple to photons, pions and most other mesons. \\
$\bullet$ It is probably considerably more compact than ordinary baryons. 

Being a flavor-SU(3) singlet, the \s\ cannot couple to flavor-octet mesons, except through an off-diagonal coupling transforming it to a much heavier flavor-octet di-baryon.   This lack of coupling to the light pseudoscalars implies a smaller spatial extent than octet mesons and baryons.  The charge radius of a nucleon is 0.9 fm in spite of its Compton wavelength being 0.2 fm, thanks to the cloud of pions surrounding it.  The \s, with no surrounding pion cloud, should have a spatial extent somewhere between its Compton wavelength, $\approx0.1$ fm, up to $\approx0.4$ fm if it has a maximal cloud of $f_0$ mesons.   

The \s-nucleon scattering cross section can be estimated to be $\sigma_{SN} \lesssim (\frac{1}{4} - 1) \sigma^{\rm el}_{NN} \approx 5-20$ mb, for $v/c \approx 1$ ($p_{\rm lab} \gtrsim 1$ GeV/c) where geometric reasoning applies.
To what extent $\sigma_{SN}$ has a low-momentum enhancement, as is the case for $\sigma^{\rm el}_{NN}$ where the cross section increases by 3 orders of magnitude as $v/c \rightarrow 0$, depends on the $SN$ potential.   The potential is difficult to predict from first principles, so for the time-being needs to be treated empirically\footnote{The $SN$ potential receives contributions from the exchange of the flavor-singlet superposition of $\omega$ and $\phi$ vector mesons (VM, mass $\sim 1$ GeV), glueballs, and the $f_0$ -- the very broad iso-singlet scalar (a.k.a $\sigma$) with mass $\sim 500$ MeV.  VM exchange produces a significant, repulsive nucleon-nucleon interaction, but \textit{a priori} the sign of the \s-VM coupling need not be the same as that of the $N$-VM coupling, so VM and f$_0$ exchange can be attractive or repulsive.  
The $f_0$ contribution to the nuclear potential is calculated in \cite{Donoghue06} using chiral perturbation theory and pion rescattering.  The method requires the parameters of chiral perturbation theory to be fixed empirically using other processes.  The $f_0$ contribution to the $SN$ potential depends on the unknown 4-point vertex $S^\dagger S\pi \pi$;  following the wavefunction overlap analysis of  \cite{fzNuc03}, this is likely much smaller than $\bar{\psi}_N \psi_N \pi \pi$. }.  The \s\ does not bind to nuclei to form exotic isotopes because the $SN$ potential is too weakly attractive or possibly not attractive at all \cite{fzBind03,fxHIDM18}.   

The production of \s\ in baryon collisions and the stability of nuclei against decay to \s\, depends on the wavefunction overlap between the \s\ and two baryons.  The overlap was calculated in \cite{fzNuc03}, for free baryons and for baryons in a nucleus, using the Isgur-Karl parameterization of the spatial distribution of quarks which allows the center-of-mass coordinate to be conveniently separated.  Fig. \ref{overlap} (adapted from Fig. 2 of \cite{fzNuc03}) shows the dimensionless overlap between \s\ and baryons ($B$s) in a nucleus, as a function of $f$, the ratio of baryon to \s\ radii, for various nuclear wavefunctions and two values of the size parameter $\alpha_B$.  
Below the red line, $|\mathcal{M}|^2_{\{B B'\} \rightarrow S} \lesssim 10^{-8}$, the formation time of an \s\ in an S=-2 hypernucleus is greater than $\sim 10^{-9}$s  (c.f. eq. (28) of \cite{fzNuc03}).   Thus non-observation of hyperon decay products \cite{ahn+hypernucBNL01,takahashi+hypernucKEK01} is comfortably compatible with the bulk of the range expected from nuclear physics.   
\begin{figure}[t]
	\label{dEdX}
	\centering
	\includegraphics[trim = 0.2in 0.0in 0.2in 0.4in, clip, width=0.48\textwidth]{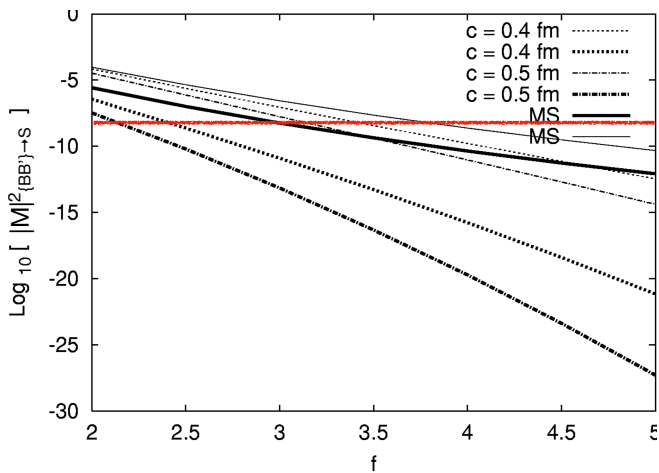}
	\vspace{-0.45in}
	\caption{Wavefunction overlap between \s\ and two baryons in a nucleus, as a function of the ratio $f=\frac{r_B}{r_S}$, for three different nuclear wavefunctions.   The heavy lines are for the standard Isgur-Karl value of the size parameter $\alpha_B = 0.406 \,{\rm fm}^{-1}$. The red fuzzy line is the approximate upper limit from doubly-strange hypernuclei.}\label{overlap}
	\vspace{-0.2in}
  \end{figure}
Such a small overlap would imply that even if $m_S < 2 m_N$, nuclei are stable on the scale of the lifetime of the Universe \cite{fzNuc03} and neutron stars do not decay to $S$'s.  (Depending on the mass and radius of the \s,  and uncertain details of the nuclear short range potential, it might be possible for SuperKamiokande and SnoLab to detect rare occurrences of nucleon fusion via processes such as $\{ p \, n \} \rightarrow S \, e^+ \, \nu_e$ or $\{ n \, n \}\rightarrow S \, \pi^0, ~ S \, \gamma$, etc., where $\{\}$ denotes {\it in a nucleus}.  Pion production might already have been noticed, suggesting it may be possible to limit $m_S$ from below to $m_S \gtrsim 2 m_N - m_\pi \approx 1.7 $ GeV~\cite{fzNuc03}.)  

Nuclei are absolutely stable and the \s\ is unstable, if $m_S > 2(m_p+m_e)$.  But  the \s\ can be effectively stable with $\tau_S > \tau_{\rm Univ}$, if $m_S  < m_p+m_e + m_\Lambda$ (so decay is doubly-weak) and the wavefunction overlap is sufficiently small.  The condition required for this was estimated to be $r_S \lesssim 0.3$ fm \cite{fzNuc03}.

How to model \s\ and \sbar\ production in hadronic interactions varies strongly with the conditions.  In exclusive or low-multiplicity interactions at relatively low energy, the overlap $|\mathcal{M}|_{B B' \rightarrow S}$ is small due to Fermi-statistics of overlapping identical baryons.  As the relative momentum of the baryons increases, the suppression of the overlap diminishes and $|\mathcal{M}|^2_{BB' \rightarrow S}$ becomes significantly larger than $|\mathcal{M}|^2_{\{B B'\} \rightarrow S}$ shown in Fig. \ref{overlap} \cite{fzNuc03}.  (A re-examination of the old H-dibaryon searches is warranted, to evaluate whether non-observation of an H with $m_H \geq 2$ GeV can be due to insufficient sensitivity, when wavefunction-overlap suppression of its production is taken into account.)

In high-multiplicity inclusive reactions,  the production rate can be estimated by extending the heuristic that there is a price of $\mathcal{O}(10^{-1})$ for each additional quark in the state, based on the meson to baryon ratio in the central region of high energy collisions and $Z$ decay.  
If applicable, this heuristic suggests an \s\ (\sbar) production rate in hadronic interactions $\approx 10^{-4}$ to $10^{-6}$ relative to pions, depending on whether a factor-10 penalty applies just for the quarks in the \s\ (\sbar) or also for those in the accompanying (anti-)baryons.  With an $\mathcal{O}(10^{-4} - 10^{-6})$ production rate relative to pions, \s's would be routinely produced in high-energy experiments.  However even at this production level they do not call enough attention to themselves to have been recognized.  The abundance of neutrons would be $\mathcal{O}(10^{3-5})$ larger than \s's, and neutrons have similar behavior:  neutral, same mass range, relatively low interaction probability in most detectors.  In cases when an \s\ or \sbar\ does interact, if noticed, the interaction would normally be dismissed as an occasional hadronic interaction of a neutron or anti-neutron.   

In addition to being electrically neutral, the \s\ has spin-0 and its spatial wavefunction is perfectly symmetric in the equal $uds$ quark mass limit.  Thus its coupling to photons is additionally suppressed by SU(3) flavor symmetry and powers of $r_S$, the \s\ radius.  For $r_S \lesssim 0.4$ fm, $S$ coupling to $\gamma$'s can only be non-neglible for momentum-transfer $\gtrsim \mathcal{ O}$(0.5 GeV), so elastic scattering of \s\ with gammas is not astrophysically or cosmologically relevant.     

\section{Strategies for discovering a stable \s\ }

Since existing methods for searching for an H-dibaryon would not have been capable of discovering a compact, stable \s\ below 2 GeV, a new approach is needed \cite{suppMat}.  The basic discovery strategy for a stable \s, if it exists, is to use the unique B= $\pm$ 2, S= $\mp$ 2 quantum numbers of the \s\ and \sbar, and the fact that B and S are conserved by the strong interactions, to infer the existence of an unseen $S$ or \sbar.  Two promising approaches are outlined below, which can in principle not only provide a smoking gun for the existence of a weak-interaction stable \s\ via ``missing BS", but also measure its mass.\\
\noindent \textit{{\bf Upsilon decay:}}  

The exclusive reactions
\be
\label{ups}
\Upsilon~~[ \rightarrow {\rm gluons}] \rightarrow S \, \bar{\Lambda} \, \bar{\Lambda}~~{\rm or} ~\bar{S} \, \Lambda \, \Lambda~~ + {\rm pions~ and/or} ~\gamma 
\ee
are ideal discovery channels.  The mass of the unseen \s\  can be reconstructed from 4-momentum conservation:  $m_S^2 = (p_\Upsilon - p_{\Lambda 1}- p_{\Lambda 2} - \Sigma p_{\pi's\&\gamma})^2$.  The width of the missing-mass peak is entirely due to resolution which is so good in some detectors, $\mathcal{O}(20)$ MeV, that even a few events appearing to be $\bar{\Lambda} \bar{\Lambda}$ or $\Lambda \Lambda $ + pions or gamma, having a common missing mass, would be a powerful smoking gun for the existence  of the \s\, and would accurately determine its mass. 

The initial state can be any $\Upsilon$ or continuum state below open-bottom threshold.  Other final states besides $\Lambda \Lambda$/$\bar{\Lambda} \bar{\Lambda}$ are also discovery avenues, e.g., $\Xi^- p$, or a $\Lambda $ can be replaced by $K^- p $.  As long as no B- and S-carrying particle escapes detection besides the \s\ or \sbar, any combination of hyperons and mesons with B= $\pm$ 2, S= $\mp$ 2 quantum numbers, including final states with higher multiplicities, can be used.   
The $\bar{\Lambda} \bar{\Lambda}$ and $\Lambda \Lambda $ final states are very good because the $\Lambda $'s short decay length ($c \tau = 8$ cm) and 64\% branching fraction to the 2-body charged final state $p \, \pi^- $, means $\Lambda$'s and $ \bar{\Lambda}$'s can be reconstructed with high efficiency, and their 4-momentum well-measured.  


At least one pair of final particles in $\Upsilon \rightarrow S \, \bar{\Lambda} \, \bar{\Lambda}$ must have $L=1$ to satisfy angular momentum and parity conservation and Fermi statistics, but the 10 GeV CM energy provides plenty of phase space.  
A high degree of hermeticity in the detector is valuable, for capturing all the particles which should be included in the missing mass reconstruction and eliminating missing baryon number and strangeness.  Even if some particles are missed, the chance of missing them in $\Delta B= \pm 2, ~ \Delta S= \mp 2$ combinations is very small and these cases would not produce a narrow missing mass peak.  Conventional background could come from missing (anti-)neutrons and $K^0_L$'s, and/or mis-IDing particles which balance the strangeness and baryon number.  The differential cross sections for conventional final states posing a spoofing challenge can be measured, so it should be possible to reliably simulate the sensitivity and expected missing mass resolution and ``dilution" from missing particles. 

A simple statistical picture described in the  Supplemental Materials \cite{suppMat} gives an estimate of the inclusive branching fraction for \s\ and \sbar\ production, BF$_S \approx 3 \times 10^{-7}$.  The Belle collaboration \cite{BelleExc13} reports branching fractions for a number of exclusive decays of $\Upsilon(1 S)$ and $\Upsilon(2 S)$,  based on samples of 102 and 158 million events respectively, with another 600M events at the $\Upsilon(3 S)$;  BaBar collected roughly 200M events on $\Upsilon(2 S)$ and $\Upsilon(3 S)$.  Thus there should be at least several hundred events with an \s\ or \sbar, and possibly many thousands, if the $10^{-4}-10^{-6}$ heuristic applies, within the existing datasets.
 

\noindent \textit{{\bf Production in hadronic collisions:}}  

Fixed target experiments in an intense beam, e.g., $K^- \, p \rightarrow S \,\bar{\Lambda}$  followed by $ \bar{\Lambda} \rightarrow \bar{p} \, \pi^+$ in NA61, can in principle discover the \s, but the small wave-function overlap between \s\ and baryons implies the rates could be negligibly small.   
A more promising strategy is to take advantage of the tremendous luminosity of the LHC, and look for characteristic decay chains after $\bar{S}$ annihilation in the beam-pipe or detector, e.g., 
\begin{eqnarray}
\label{Sbarannih}
\bar{S}+N &\rightarrow& \bar{\Xi}^{+,0} + X, \,{\rm with}\,\, \bar{\Xi}^{+,0} \rightarrow \bar{\Lambda} \pi^{+,0} \,\&\, \bar{\Lambda} \rightarrow \bar{p} \pi^+~  \nonumber \\{\rm or}~
\bar{S}+N &\rightarrow& \bar{\Lambda} + K^{+,0} + X.
\end{eqnarray}
$\bar{S}$'s should have a similar transverse-energy distribution as other hadrons, i.e., $<p_t >\lesssim \mathcal{O}(1)$ GeV.  The anti-baryon produced in $\bar{S}$ annihilation in a central tracker will commonly be a $\bar{\Xi}^{+,0}$ with $\gamma \sim 1$ which decays in $\mathcal{O}(5)$ cm to $\bar{\Lambda} \pi^{+,0}$, followed 64\% of the time by $\bar{\Lambda} \rightarrow \bar{p} \pi^+$ in $\sim$ 8 cm.  Observing such distinctive production/decay chains provides unambiguous evidence for a B=-2, S= +2 neutral particle that initiated the annihilation interaction in the beam pipe or detector.  Complementing $\bar{\Xi}$ production are $\bar{S} n \rightarrow \bar{\Lambda} K^0_S$, where the double ``$V$"s locate the annihilation vertex and $\bar{\Lambda}$ ID proves the initiating particle has $B=-2$, and $\bar{S} p \rightarrow \bar{\Lambda} K^+$, where the $B=-2, S=+2$ of the initial state is unambiguous.

There are $\approx 30$ charged particles with pseudo-rapidity $|\eta | < 2.4$, for 7 TeV LHC  p-p collisions \cite{CMSchgpartmult10}, so the number of \sbar's produced with pseudo-rapidity $|\eta | < 2.4$ in a dataset with $N_{11}10^{11}$ recorded interactions is  
$
N_{\bar{S}} \approx  3 f^{\rm prod}_{-4} \, N_{11} \, 10^{8},
$
where $f^{\rm prod}_{-4} 10^{-4}$ is the \sbar\ production rate relative to all charged particles in the given rapidity range.  The material budget for the CMS tracker and beam pipe ranges from 0.12-0.55 hadronic interaction lengths in this $\eta$ range; take 0.33 for an estimate.  Write the \sbar\ annihilation cross section as $\sigma_{\bar{S}N} \equiv f^{\rm annih}_{-6} 10^{-6} \sigma_{NN} $, acknowledging that the annihilation reaction is at relatively low energy so the penalty of the overlap suppression is more severe than that for production in the original high energy collision.  Finally, take the fraction of annihilation final states containing $\bar{\Xi},\bar{\Lambda}$ to be $f_{\bar{\Xi},\bar{\Lambda}}$, where we expect $f_{\bar{\Xi},\bar{\Lambda}} \sim \mathcal{O}$(1) for low $\sqrt{s}$ annihilation.  Thus the number of potentially reconstructable annihilation+decay chains is 
\be
 N_{\bar{\Xi},\bar{\Lambda}} \, = \, f^{\rm prod}_{-4} \, f^{\rm annih}_{-6} \, f_{\bar{\Xi},\bar{\Lambda}} \, N_{11} ~~ 10^{5}.
\ee
This is encouraging, and if using the existing dataset proves insufficient, the distinctive decay chains may allow for a dedicated trigger which could effectively increase the sample by up to a factor $\approx 10^5$, for any given integrated luminosity. 
 
In some subset of events, all the final particles of the annihilation will be identified and their 3-momenta adequately measured.  With these events, the 3-momentum and kinetic energy of the \sbar\ can be deduced from energy-momentum conservation, modulo the nuclear Fermi-momentum of the nucleon on which the \sbar\ annihilates.  In principle, this should enable the mass of the \sbar\ to be measured.  Whether this is feasible in practice must be determined empirically and with detailed detector simulations.  Production of \s\ and \sbar\ has been implemented in the microcanonical fireball contribution to final states in the event generator EPOS-LHC, which will facilitate detector studies for LHC and for $\Upsilon$ decay \cite{fpw18}.

\section{Conclusion}   

Existing experiments and current knowledge of non-perturbative QCD are compatible with the existence of a so-far-undiscovered stable or very long-lived B=2, S=-2, Q=0 sexaquark denoted \s.   Past experiments probably rule out $m_S\gtrsim 2$ GeV, but the crucial range below 2 GeV has yet to be explored.  

Unambiguously demonstrating the existence of a population of events with apparent baryon-number and strangeness violation, in the pattern $ \Delta B=\pm 2,~\Delta S = \mp 2$, 
would be a smoking-gun for the existence of the \s.  However due to the small wave-function overlap between \s\ and two baryons, the production rate of \s's is very small in low energy hadronic collisions, whereas in high energy collisions, with higher production rate, the \s\ is difficult to identify due to its similarity to the vastly more copious neutrons.  

Two experimental strategies to discover a stable \s, which overcome these problems, have been proposed here.  In both cases, the \s\ or \sbar\ itself is not directly observed:
\\
$\bullet$  Search in $\Upsilon$ decays for final states with apparent B=$\pm 2$ and  S = $\mp2$, corresponding to, e.g.,  $\bar{S} {\Lambda} {\Lambda}$ or $S \bar{\Lambda} \bar{\Lambda}$ (plus pions). \\
$\bullet$  Search in LHC events for the B=-1, S=+2 signature of the final state of $\bar{S}$ annihilation in the beam pipe or detector.  

The two new proposed strategies overcome the problems with a conventional search in complementary ways:   in $\Upsilon$ decay, the short-distance nature of the gluonic source reduces the penalty associated with the \s's spatially compact wave function, while the LHC search for $\bar{S}$ annihilation in the beam pipe and detector, exploits a distinctive signature which makes finding the needle-in-the-haystack feasible.  

Belle and Babar 
may have hundreds or thousands of events containing \s\ and \sbar's, and the LHC could have even more events with \sbar\ annihilating in the beam pipe or detector.  
The mass of the \s\ should be measurable via missing mass in exclusive $\Upsilon$ decay, and can in principle be deduced from energy-momentum conservation in LHC \sbar\ annihilation and subsequent decay chains.   
  
\noindent{\bf Acknowledgements:}
I wish to thank  J. D. Bjorken, J. Donoghue and N. Wintergerst for discussions about the \s\ and modeling its interactions with other hadrons,  S. Beane, G. P. LePage and M. Savage for discussions about lattice QCD efforts to measure the mass of the \s/H-dibaryon,  L. Littenberg for information and discussions about previous H-dibaryon searches, and B. Echenard, R. Musso, S. Olsen,  R. Ulrich and M. Unger for information regarding the capabilities of various current detectors.  This research was supported by NSF-PHY-1212538 and the James Simons Foundation.  The hospitality of the SLAC Particle Theory Group and the Kavli Institute for Particle Astrophysics and Cosmology during part of this work, is gratefully acknowledged.  

\def\apj{Astrophys.\ J.}
\def\nat{Nature}
\def\apjl{Astrophys.\ J. Lett.}
\def\apj{Astrophys.\ J.}
\def\aap{Astron.\ Astrophys.}
\def\prd{Phys. Rev. D}
\def\physrep{Phys.\ Rep.}
\def\mnras{Month. Not. RAS }
\def\araa{Annual Rev. Astron. \& Astrophys.}
\def\aapr{Astron. \& Astrophys. Rev.}
\def\aj{Astronom. J.}
\def\jcap{JCAP}



%

\newpage

\section{ Supplemental Materials} 
\subsection{Statistical model for \s/\sbar\ production in $\Upsilon$ decay}

All $\Upsilon$ decays below open-bottom, i.e., $\Upsilon(1S)$, $\Upsilon(2S)$ and $\Upsilon(3S)$ decays, go through 3-gluons.  To leading order each gluon converts to $q$ and $\bar{q}$s of  $\approx 2$ GeV, which produce mini-jets or form hadrons through string-fragmentation.   Production of the minimum 6$q+6 \bar{q}$ with 6 $q$'s or $\bar{q}$'s having relatively similar momenta needed to produce an \s\ or \sbar\, requires creating 3 more gluons, at a penalty factor of $\alpha_s^3$ in the amplitude.  Here $\alpha_s \gtrsim \alpha_s(\Upsilon) \approx 0.2$, and may be $\mathcal{O}(1)$ because large momentum transfer is not required, so we adopt the geometric mean.    

The next requirement is for 6 $q$ or $\bar{q}$'s to be nearest neighbors in space, within a distance scale $<r_S$.  Statistically, the penalty for having exclusively $q$'s or $\bar{q}$'s within a nearest-neighbor grouping is $\left( \frac{1}{2} \right)^5$.  As a zeroth approximation, no penalty is included for such a grouping of $q$ or $\bar{q}$'s to have an appropriate spatial wavefunction to be an \s\ or \sbar, on the grounds that the $q$ and $\bar{q}$'s originate in a region of size $\approx \frac{1}{10\rm GeV} = 0.02$ fm then expand, so at some point they will be in the relevant volume to form an S. 

Finally, to form an \s\ or \sbar, the 6 $q$'s or $\bar{q}$'s must have the total flavor-spin-color quantum numbers to be an \s\ or \sbar.   Without loss of generality consider the 6$q$ case.  \s\ belongs to the (1,1,1) representation of $SU(3)_c \times SU(3)_f \times SU(2)_s$.  With a spatially symmetric wavefunction, Fermi statistics implies it is in the totally antisymmetric 6-quark representation of $SU(18)$.   In a statistical approximation that the $q$'s produced by the gluons randomly populate all possible color, flavor and spin states, the fraction of cases contributing to the antisymmetric representation is $18\times17\times16\times15\times14\times13/(6\times5\times4\times3\times2)/18^6 =  5.46 \times 10^{-4}$.  Dividing by $2^5$, multiplying by $(\alpha_s^2 \approx 0.2)^3$ and by 2 for \s\ and \sbar\, results in the estimated branching fraction of Upsilon to states containing an \s\ or \sbar:  BF$_S \approx 2.7 \times 10^{-7}$.  

A more detailed model calculation could give a larger estimate, since color correlations among the gluons and quarks likely enhance the amplitude for states corresponding to attractive QCD channels, while the simplification of ignoring the spatial structure could lead to a suppression.  Combining these effects with the crude nature of the estimated dependence on $\alpha_s$,  means improvements can take this naive estimate in either direction.  

\subsection{Experiments excluding a long-lived loosely-bound $H$ dibaryon}
The BNL E888 collaboration performed two different searches for the $H$ dibaryon~\cite{belz+96} and \cite{belz+DiffDissocH96}.  The latter uniquely among existing experiments, did not restrict its search to mass greater than 2 GeV or to decaying $H$ dibaryon.   A 24.1 GeV/c proton beam from the Brookhaven National Laboratory AGS, was directed onto a Pt target from which a neutral beam cleaned of photons was produced at 65 mr.  Approximately 18 m down stream, it struck a scintillation counter system, ``the dissociator", which was followed by drift chambers, magnet, trigger counters and Cerenkov counter, optimized for identifying diffractively produced $\Lambda$'s.  A total of $4 \times 10^7$ events were recorded. They placed a limit on the product of cross sections for $H$ production and dissociation:
\be
\frac{d \sigma_H}{d \Omega}|_{\rm 65\, mr} < \left( \frac{N^H_{\Lambda \Lambda} A_{\Lambda K_S} \sigma^c_{\Lambda K_S}}{N^c_{\Lambda K_S} A_{\Lambda \Lambda} \sigma_{\Lambda \Lambda}}\right) ~ \frac{d \sigma_n}{d \Omega}|_{\rm 65\, mr}~,
\ee
where $N^c_{\Lambda K_S}$ is the number of coherently produced $\Lambda^0 K^0_S$, $A_{\Lambda \Lambda}$ and $A_{\Lambda K_S} $ are the acceptances and efficiencies for $H + A \rightarrow \Lambda^0 \Lambda^0 A \rightarrow p \pi^- p \pi^- A$ and $ n + A \rightarrow \Lambda^0 K^0_S X \rightarrow p \pi^- \pi^+ \pi^- X$ respectively, where diffractively dissociated final states are characterized by the distinctively forward-peaked distribution, and $\sigma_{\Lambda \Lambda}$ and $\sigma^c_{\Lambda K_S}$ are the respective diffractive dissociation cross sections.   Using their acceptance estimates, upper limits on $N^H_{\Lambda \Lambda}$ taking into account estimates of backgrounds, and knowledge of  $\frac{d \sigma_n}{d \Omega}$, they find for 24.1 GeV/c p-Pt collisions at 65 mr:
\be
\frac{d \sigma_H}{d \Omega}|_{\rm 65\, mr} ~ \frac{\sigma_{\Lambda \Lambda}}{0.5 mb} < 2.3\times 10^{-4}   \frac{d \sigma_n}{d \Omega}|_{\rm 65\, mr}~\frac{\sigma^c_{\Lambda K_S}}{5.9 \mu b} ~.
\ee

Although the conventional H-dibaryon scenario, in which the H is a relatively loosely bound state of two $\Lambda$'s is  constrained by these limits, this limit is far too weak to be constraining for the tightly bound \s\ scenario.  The diffractive dissociation cross section has a wavefunction overlap penalty of $|\mathcal{M}|^2_{H \rightarrow {\Lambda\Lambda} }$; as discussed in the main text, this is $\lesssim 10^{-8}$.  A similar wavefunction overlap penalty also is expected for producing $H$'s in the absence of the conditions discussed in the main text, e.g., the small source size in $\Upsilon$ decay.

For a review of the eight other H-dibaryon searches, see \cite{chrienHBNL98}.   We also remark that the claim of \cite{ejiri+H89} that a loosely bound H with mass more than a few MeV below $m_p + m_n$ can be ruled out, was shown \cite{fzNuc03} to overlook the crucial role of the wavefunction overlap in any stable sexaquark scenario.

\subsection{Naming}
Greek prefixes have been used for 4- and 5-quark states expanding on mesons and baryons with an additional $q \bar{q}$ pair, thus ``hexa" should be reserved for a possible $q \bar{q} q \bar{q} q \bar{q}$ state and we seek a Latinate prefix.  ``Six" as in ``six-quarks" is a cardinal number, and as can be seen from the Wikipedia table at https://en.wikipedia.org/wiki/Numeral\_prefix , the correct Latinate prefix for cardinal-6 is ``sexa".   ``Sext" or ``sexta" is appropriate when the prefix relates to an ordinal number.  (In certain words, such as quartet, quintet, sextet... either the origin evoked an ordinal number or the ``t" was added for smoothness.)   

\end{document}